# Experimental Generation of Intrinsic Localized Modes in a discrete electrical transmission line

Ryan Stearrett and L.Q. English
*Department of Physics and Astronomy, Dickinson College, Carlisle, PA 17013, USA*

Experimental evidence for the generation of intrinsic localized modes (ILMs) in a nonlinear electrical transmission line is presented both via modulational instability (MI) of the uniform mode and via driving the lattice locally. The spatial profiles of these modes localized on the order of the lattice spacing can be directly measured in this macroscopic lattice, in contrast to most other systems where ILMs have so far been detected.

## 1. INTRODUCTION

Intrinsic localized modes (ILMs) are stable, nonlinear excitations in perfectly periodic lattices which are spatially localized on the order of a few lattice constants. ILMs, also called discrete breathers or lattice solitons, have attracted enormous attention in many areas of physics since they were first postulated in the late 1980s [1, 2], and they have been extensively studied in theoretical, computational and (more recently) experimental settings [3]. Historically, the small class of integrable nonlinear lattices, such as the Toda lattice [4] was tackled first, where exact ILM solutions could be obtained analytically. Soon thereafter, ILMs were found numerically in more realistic atomic lattice models [4-7], and since then, the class of systems for which ILMs have been discovered numerically has greatly expanded.

On the experimental side, efforts can generally be classified according to the scale of the system. In nanoscale systems, ranging from antiferromagnetic spin-lattices [9, 10] to atomic vibrations in bcc $^4$He [11], spectroscopic evidence of ILMs has been reported but must necessarily remain somewhat indirect. In mesoscopic systems, such as arrays of Josephson junctions [12, 13] and micromechanical cantilever arrays [14, 15], ILMs have been seen more directly with some spatial resolution. Even here, however, a time-resolved image of the ILM profile cannot be achieved.

In this paper, we present experimental evidence of ILMs in a macroscopic lattice – a discrete electrical transmission line. The advantage of this system is that its spatial scale allows for the detailed characterization of ILM profiles, which is usually only possible in numerical simulations. Its resonance frequencies (of a few hundred kilohertz) are comparable to the micromechanical oscillator system.

Conceptually, ILMs are akin to impurity modes in crystals with defects. It is well known that an impurity can act as a localization center, giving rise to modes that fall outside of the spectrum of extended solutions. In pure lattices, an effective impurity can be self-generated in a region of large oscillation amplitude via the nonlinearity of the medium. Loosely speaking, the ILM creates its own effective impurity in a self-consistent way, relying on the nonlinearity of the lattice. The ILM is also conceptually related to the soliton, which is a nonlinear localized excitation in a continuous medium, with the important distinction that ILMs exists and depend upon the discreteness of the lattice. For this reason, ILMs are sometimes referred to as lattice solitons.

Generally speaking, ILMs have proven to be so ubiquitous precisely because they are fairly generic excitations in nonlinear, discrete systems. It is plausible, therefore, to expect the electrical lattice of this study to support ILMs as well, in particular because it exhibits similar dispersive and nonlinear characteristics as other systems in which they have been detected. Indeed, localized excitations have recently been theoretically predicted as solutions to the discrete nonlinear Schrödinger equation to which the equations of motion governing the electrical transmission line can be approximated [16].

Previous studies of discrete electrical transmission lines [17-20] have focused on traveling solitons generated from a pulse inputted at either end of the transmission line. For single-inductance lattices, the pulse was shown to deform into the characteristic soliton profile as it traveled along the line, and for bi-inductance lattices, envelope solitons were similarly generated.

This paper focuses on sharply localized, non- or slowly propagating ILMs, instead of the broader, traveling solitons. For this purpose, the geometry is modified to a ring structure with no boundaries. First, we consider a homogeneous driver initiating ILMs via modulational instability (MI) of the uniform mode [21]; thereafter, the effect of local driving and direct coupling to ILMs is investigated.

## 2. EXPERIMENTAL DETAILS

A schematic diagram of the bi-inductance transmission line used in these experiments is shown in Fig.1. Each unit cell of this lattice consists of two inductors, $L_1$=680μH and $L_2$=330μH, a varactor diode (NTE 618) characterized by a voltage-dependent capacitance $C$, and a DC-blocking capacitor of a comparatively large capacitance (1μF). Each unit-cell or node can be driven via a resistor, $R$=10 kΩ, by a sinusoidal signal with the possibility of a DC offset for reverse-biasing the diode. The capacitance of the varactor diode is found to be 450pF at 1V bias, and it decreased to 140pF at 4V. Even at no bias voltage, the effective capacitance is around 800pF. It is this dependence of the capacitance on voltage that imparts nonlinearity to the circuit. It should be noted that the resistors are technically not part of the circuit to be investigated, but are necessary for driving the circuit.

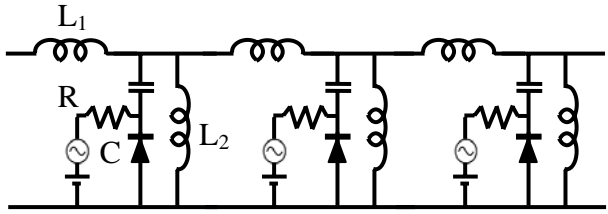

Figure 1 :   Schematic circuit diagram of the electrical transmission line.

The low-amplitude plane-wave dispersion curve for the bi-inductance lattice is given by

$$\omega^2 = \frac{1}{L_2 C} + 4\left(\frac{1}{L_1 C}\right)\sin^2\frac{\kappa}{2}, \qquad (1)$$

where $\kappa$ is the wavenumber of the voltage wave. Thus, the zero-bias frequencies of the uniform mode ($\kappa=0$) and zone boundary mode ($\kappa=\pi$) are computed as 320 kHz and 540 kHz, respectively, in good agreement with experimental data. Note that the second harmonic of the uniform mode is above the zone boundary and thus falls outside the plane-wave spectrum. If this were not the case, the uniform mode could not undergo MI and ILMs – appearing below the uniform mode - would not be stable excitations.

Equation 1 yields the familiar S-shaped dispersion curve with a gap below the uniform mode. For small wave numbers, $\kappa$, the sign of the curvature of the dispersion curve and that of the nonlinear coefficient render MI possible in this region.

In contrast to previous studies on nonlinear bi-inductance transmission lines, we consider here a closed lattice of 32 unit cells. The geometry of the lattice is a ring, which eliminates the boundaries and makes all nodes equivalent. The system can then be driven as a whole, rather than inputting a signal at the first node of a line-circuit and observing it propagate to the last node [18].

The basic experimental setup is illustrated in Fig. 2. The sweep generator (BK Prec 4017A) serves as the sinusoidal driver of the system; its frequency is controllable via a voltage input. Thus, in order to take spectra, we incorporate a second function generator with a triangular voltage output. The response at a particular node can then be displayed against frequency on an oscilloscope, if it is set to XY display mode and the y-axis turned to envelope display. To stay at a particular frequency, we flip the switch and connect a DC power supply to the sweep generator.

The ring circuit was built on a standard breadboard, and the driving signal is inserted along a vertically connected strip on the breadboard into which each unit cell can tap. When driving the ring circuit in this manner, the spectrum obtained is independent of the node that is being probed, demonstrating the equivalence of all nodes. The voltage probes do not introduce an impurity into the circuit since they have high input impedance and low capacitance.

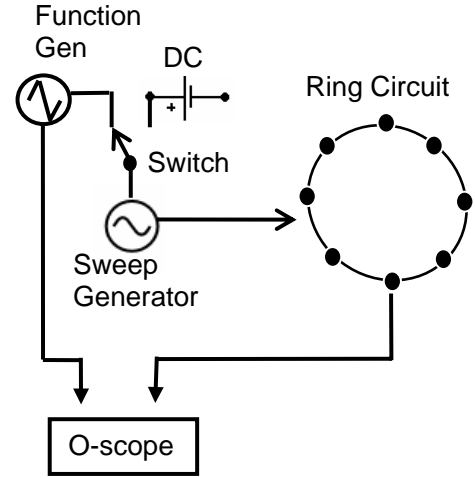

Figure 2 :   The basic experimental setup.

## 3. RESULTS AND DISCUSSION

We start by examining the uniform mode excitation supported by the lattice. This is best done by driving the system uniformly with all nodes connected to the sweep generator, as explained in the last section. In this way, the only mode visible is the uniform mode, the driver being unable to couple to the rest of the plane-wave spectrum. Figure 3 shows the results for three different driver amplitudes. At low amplitudes, we observe a fairly symmetric response peak centered at 325 kHz. The width of the peak indicates that damping due to energy dissipation is significant in our circuit, yielding a Q-factor for this resonance of around 10. The source of the dissipation is most likely the ohmic resistance of the inductors of around $2\Omega$.

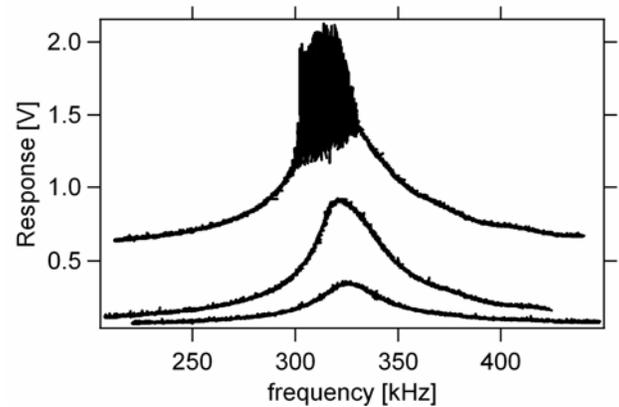

Figure 3 :   Spectra showing the uniform mode at various driving amplitudes. At large amplitudes, modulational instability is observed (dark region).

As the driving amplitude is increased, the response curve is seen to bend towards the left, as expected for a system characterized by soft-nonlinearity. Note that all traces shown here represent increasing frequency sweeps and that for the intermediate trace, slight hysteresis can already be observed when comparing up- and down- sweeps. At the largest driving amplitude in Fig.3, the response amplitude has crossed the MI threshold, as indicated by the black region in the up-most trace. Here the response is not single-valued but varies rapidly over a large voltage interval.

It is well known that the initial MI of a plane-wave mode can give rise to fully nonlinear localized modes at longer times in a wide variety of physical systems [22-25]. In order to investigate whether localized modes are produced by the MI in this system, we probe the lattice at four successive nodes simultaneously, while driving the lattice uniformly at large amplitude.

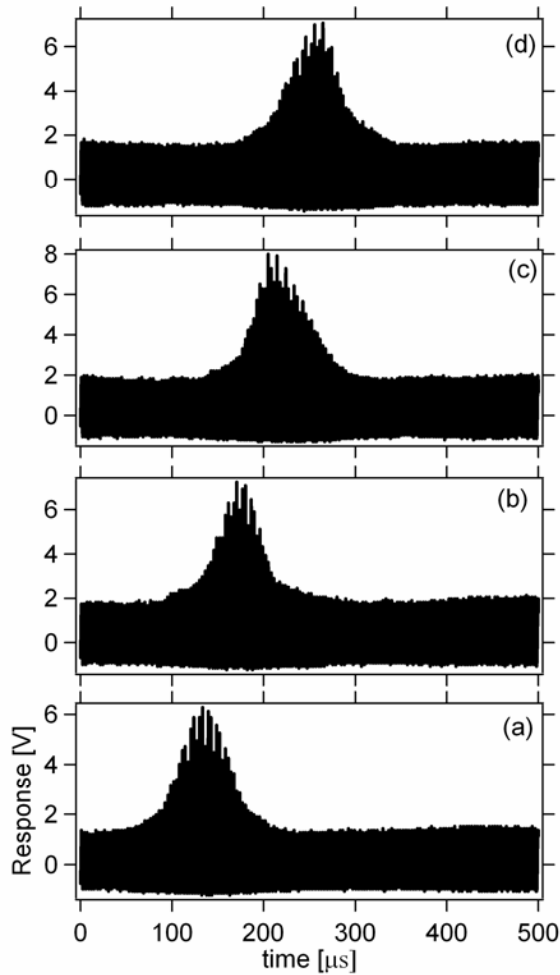

Figure 4 :   A localized excitation traveling to the right. The individual panels display the time-dependent response at four neighboring nodes.

The amplitude and frequency of the driver are adjusted to place the uniform mode in the modulationally unstable region. The oscilloscope was triggered by the driving signal. Figure 4 shows the data from the four successive nodes in the lattice. Note that the x-axis measures time in microseconds. Localization of energy has clearly been initiated in Fig. 4 (a), where the peak response occurs at around 125 µs. At the next three nodes, Figs. 4 (b)-(d), this peak shifts to larger times, indicating that the localized mode is traveling from left to right. The speed of this ILM in these frames is on the order of 0.02 cell/µs, which is about 50 times less than the speed reported for envelope solitons [20].

To ascertain the spatial extent of this particular excitation, Fig. 5 expand the time axis to zoom in on a few oscillations at the neighboring nodes. The center node attains a peak voltage of 8V at 7.4 µs, whereas the nearest neighbors reach only about 3V, and the next-nearest neighbor (lowest trace) has decreased to 1.6V at that time.

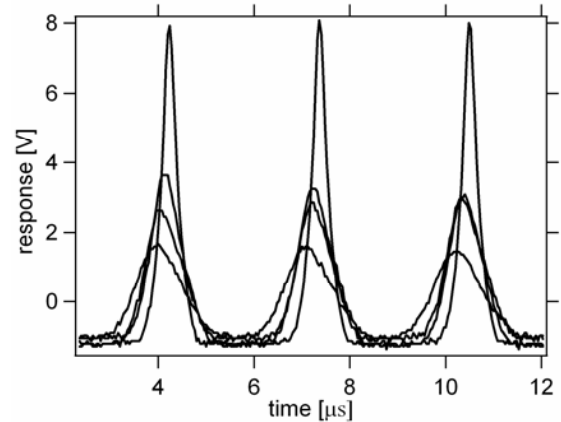

Figure 5 :   Similar to the previous figure, but with the time-axis expanded. By comparing values at a particular time, the spatial profile is ascertained.

We conclude that at a given instant in time, the spatial profile is localized on the order of the lattice spacing. This observation justifies calling the excitation an ILM or discrete breather as compared to a soliton.

One characteristic of these ILMs produced via MI of the uniform mode is that they are numerous in the lattice and hop fairly randomly from node to node. The role of the homogenous driver, it seems, is mainly to perpetuate the MI. Since the lattice can evidently support ILMs, the question arises of whether a local driver can directly couple to them. Thus, we examine the effects of driving the lattice at one particular (but arbitrary) node. The sweep generator of output impedance 50 Ω is introduced into the circuit via a large (5 kΩ) resistor. The 10 kΩ resistors connected to the other nodes are - in the absence of the driver - connected directly to ground and could potentially be removed altogether. In the results shown here, we have left the 10 kΩ resistors in place, but we obtain qualitatively similar results when they were removed.

With this local driving in place, all the plane-wave modes consistent with the periodic boundary conditions should be visible in the spectrum even in the linear driving regime, and indeed when the driving frequency is swept over a large frequency interval, we now see a multitude of modes starting with the uniform mode at 312 kHz and extending to the zone boundary at 535 kHz. For this reason, the discrete transmission line can be used as a band-pass filter.

Figure 6 depicts the various spectra obtained at increasing amplitudes of the local driver. The lower panel, Fig. 6 (a), displays the lower three amplitudes, whereas the upper panel, Fig. 6 (b), shows the response at the largest driving amplitude for scanning up in frequency (solid line) and scanning down (dotted line).

Starting with the lower panel, it is evident that the uniform mode is characterized by soft-nonlinearity as in the case of homogenous driving, bending to the left on the frequency axis and finally becoming unstable against modulations. At the high-frequency end of the spectrum, one noteworthy feature is the sudden appearance of a peak/shoulder around 620 kHz in the third trace of Fig. 6(a).

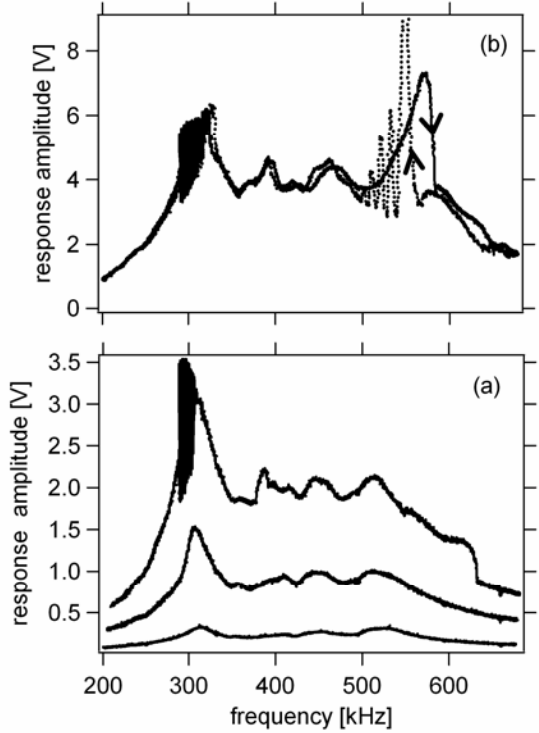

Figure 6 : Spectrum obtained from a local driver. (a) Spectra of increasing driver amplitude. MI appears in the third trace. (b) For largest driving amplitude, large hysteresis. Solid line: up-sweep, dotted line: down-sweep.

We interpret it as the parametric resonance of the uniform mode at twice its frequency, which is borne out by spatial profile studies. In Fig. 6 (b), we can clearly detect hysteresis near the zone boundary and - to a lesser degree - at the zone center.

Next, we set the driving frequency to 294 kHz, i.e. to the left of the linear uniform mode resonance and in the unstable region. The highest attainable amplitude is selected, corresponding to the traces in Fig. 6(b). Figure 7 shows two time snapshots of the voltage response at 13 neighboring lattice nodes, with node 6 connected via the resistor to the driver. The excitation is clearly seen as localized in space, extending no farther than about three or four lattice constants in either direction. By the fifth node away from the center, the signal has almost completely vanished.

Since there is some dissipation in this lattice, we would expect the response to a local driver to gradually diminish with distance from the driving node. Thus, we need to demonstrate that the localized nature of the excitation at 294 kHz cannot be explained by dissipation but is truly a nonlinear effect. For this purpose, we now compare the node-by-node response as a function of time of two identifiable modes in the spectrum, each driven at the highest amplitude.

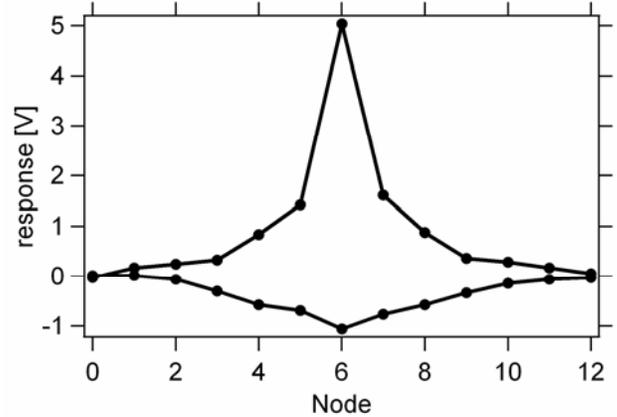

Figure 7 : Spatial profiles of the ILM in the frequency gap at two different times illustrating the minimum and maximum extent of the ILM's motion.

Figure 8 depicts the resulting density plot. The x-axis displays time and the y-axis node number. The color indicates response amplitude, with black corresponding to the largest and white the smallest voltage. Both panels in the figure use the identical voltage-to-color scheme.

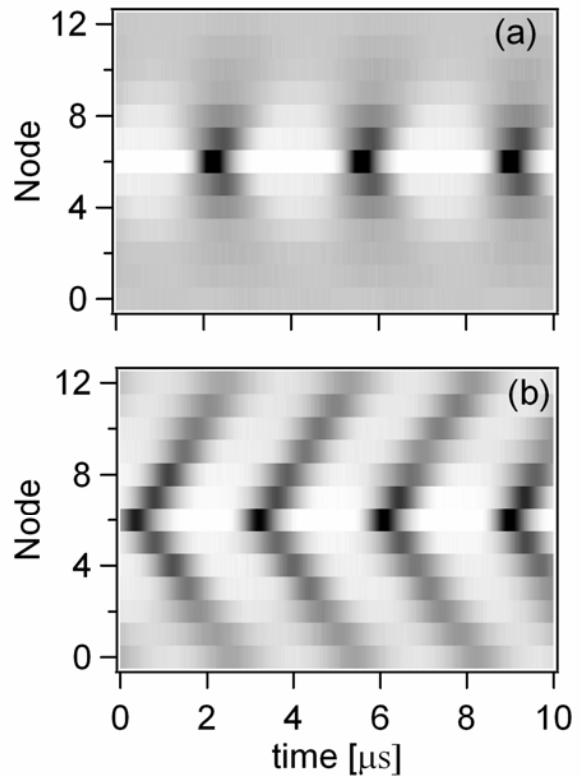

Figure 8 : Density plots illustrating the ILM and plane-wave signatures in space (y-axis) and time (x-axis). Color indicates node voltage, black representing high voltage. (a)

localized mode at 294 kHz, (b) plane-wave mode at 351 kHz.

The upper panel depicts the ILM whose profile is given in Fig. 7. The energy is clearly localized at the 6$^{th}$ node; interestingly, at that node, the energy is also periodically localizing in time. We compare this distribution of energy to that of the lower panel in Fig. 8 (b). There the frequency is set to 351 kHz which places the driver inside the dispersion curve, thus exciting a plane-wave mode to large amplitudes. Indeed, we notice that the signals from the various nodes are no longer in phase with one another, as expected for a non-zero wavenumber. The important observation, however, is that the response amplitude does not exhibit the sharp localization evident in the upper panel. Instead, the signal strength decreases much more gradually due to the dissipation in the lattice.

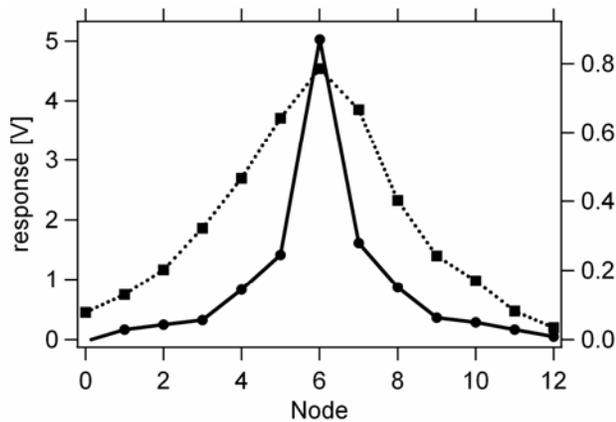

Figure 9 : Comparison of the response versus node number at large driving amplitude (solid line) and at low amplitude (dotted line). The emergence of an ILM at lower frequency is observed.

Alternatively, it is possible to compare the spatial profiles of the ILM with that of the low-amplitude uniform mode. In Fig. 9, the dotted line represents the node-by-node response at low driving amplitudes (see right axis) corresponding to the uniform mode. Again, the voltage gradually diminishes due to ohmic losses. When the amplitude is significantly increased (see left axis), the spatial response is considerably more localized about the central node. Figures 8 and 9 present data that, when taken together, point quite conclusively to nonlinear self-localization in this system.

## 4. CONCLUSIONS

In this paper, we have experimentally verified the existence of ILMs in a nonlinear electrical transmission line for the first time and directly mapped out the spatial and temporal profile of the ILM solution. We have shown that ILMs can be produced by a homogeneous driver via the MI of the uniform mode which spontaneously emerges beyond a particular amplitude threshold. ILMs can also be created more directly via a local driver which can couple to and excite these localized modes.